\documentclass[a4paper]{article}

\usepackage{INTERSPEECH2022}
\usepackage{amsthm,amsmath,amssymb}
\usepackage{multirow}
\usepackage{enumerate}
\usepackage{makecell}
\usepackage{cite}

\usepackage{mathrsfs}

\title{Back-ends Selection for Deep Speaker Embeddings}

\name{Zhuo Li, Runqiu Xiao, Zihan Zhang, Zhenduo Zhao, Wenchao Wang, Pengyuan Zhang}
\address{
 Key Laboratory of Speech Acoustics and Content Understanding, Institute of Acoustics, Chinese Academy of Sciences, Beijing, China\\
 University of Chinese Academy of Sciences, Beijing, China
}
\email{\{lizhuo,xiaorunqiu\}@hccl.ioa.ac.cn}

\begin{document}

\maketitle
\begin{abstract}
Probabilistic Linear Discriminant Analysis (PLDA) was the dominant and necessary back-end for early speaker recognition approaches, like i-vector and x-vector. However, with the development of neural networks and margin-based loss functions, we can obtain deep speaker embeddings (DSEs), which have advantages of increased inter-class separation and smaller intra-class distances. In this case, PLDA seems unnecessary or even counterproductive for the discriminative embeddings, and cosine similarity scoring (Cos) achieves better performance than PLDA in some situations. 
Motivated by this, in this paper, we systematically explore how to select back-ends (Cos or PLDA) for deep speaker embeddings to achieve better performance in different situations. 
By analyzing PLDA and the properties of DSEs extracted from models with different numbers of segment-level layers, we make the conjecture that Cos is better in same-domain situations and PLDA is better in cross-domain situations. We conduct experiments on VoxCeleb and NIST SRE datasets in four application situations, single-/multi-domain training and same-/cross-domain test, to validate our conjecture and briefly explain why back-ends adaption algorithms work.

\end{abstract}
\noindent\textbf{Index Terms}: speaker verification, PLDA, Cosine similarity scoring, domain adaption

\section{Introduction}

Speaker recognition aims to verify the identities of speakers from samples of their voices, which has been deployed in many commercial products successfully. 
In the early years, i-vector\cite{ivector} front-end and PLDA\cite{plda,plda02} back-end are the dominant model since they are proposed.
With the rise of neural networks, discriminatively trained DNNs or CNNs\cite{xvectorbefore,snyder2019speaker,liming,resnet02} surpass i-vector as the state-of-the-art front-ends. Extensive work focuses on improving discrimination of front-ends, developing more comprehensive neural architectures\cite{res2net,ecapa}, improving pooling methods\cite{okabe2018attentive,tang2019deep,liming,wang2020multi}, or exploring effective objective functions\cite{liu2019large,zhou2020dynamic,8759958,xiao21b_interspeech}. 

Two phenomena have attracted our attention with significant advances in the discrimination of front-ends.
Firstly, Cos, which is simple, achieves better performance than PLDA in certain situations, and PLDA gradually loses its dominance.
In a series of VoxSRC\cite{1fcetdnn,thienpondt2020idlab,wang2020dku,zhang2021beijing,brummerbut,wang2021xmuspeech}, almost all participating teams use Cos rather than PLDA. 
PLDA is usually adopted in the previous SREs\cite{lee2019i4u,lee2020nec,villalba2019state,li2020thuee,li2020voiceai,alam2019abc,ramoji2020leap}, whereas half of the participating teams use Cos instead of PLDA in the recent NIST SRE21\cite{sadjadi2021nist,avdeeva2021stc}.
Thus, we wonder how to select back-ends for deep speaker embeddings to obtain better performance in various situations.

Secondly, it is noticed that the number of fully-connected layers in the segment-level part differs across studies, with some being one\cite{1fcetdnn,thienpondt2020idlab,wang2020dku,zhang2021beijing,sadjadi2021nist} and some being two\cite{lee2019i4u,lee2020nec,villalba2019state,li2020thuee,li2020voiceai,alam2019abc,ramoji2020leap}, and embeddings distributions vary among layers. Some studies \cite{9023015,9296778} argue that the distribution of DSEs is the major reason why PLDA does not work on DSEs. They constrain distributions to alleviate this problem, such as VAE\cite{9023015}, deep normalization\cite{9023015}, etc.
There, we wonder whether the number of segment-level layers affects back-ends selection.

In the paper, by analyzing PLDA and the simplified PLDA log-likelihood ratio (LLR) in detail, we find that Gaussian assumption and utilizing the training data distribution are two major differences between PLDA and Cos.
It is noticed that strong discrimination and non-Gaussianity are two major properties of DSEs, which are better for Cos.
Because the Gaussian assumption is one of the most underlying assumptions of PLDA, non-Gaussianity of DSEs limits the performance of PLDA. 
While, utilizing the training data distribution leads PLDA to be more robust than Cos when encountering cross-domain problems. The variance of PLDA represents the discrimination in each diminsion. PLDA assign different weights to different dimensions according to their variance to maintain robust in cross-domain situations.
According to these, we conjecture that Cos is better in same-domain situations and PLDA is better in cross-domain situations. In addition, discrimination and non-Gaussianity of DSEs vary across the number of fully-connected (\textit{fc}) layers in the segment-level part, which further affects back-ends selection.

Finally, we conduct experiments on VoxCeleb and NIST SRE datasets in four application situations, single-/multi-domain training and same-/cross-domain test. Our experiments results validate our conjecture. 
In addition, we briefly conduct some validation and discussion on domain adaption algorithms.

\section{PLDA and Deep Speaker Embeddings}
\subsection{PLDA}
\label{sec211}
\subsubsection{Revisting PLDA}
we consider the two-covariance form of the PLDA model:
\vspace{-1.5mm}
\begin{equation}
\begin{aligned}
P(x|y)=\mathcal{N}(x|y,\Phi_w) \quad P(y)=\mathcal{N}(y|m,\Phi_b)
\end{aligned}
\vspace{-1.5mm}
\end{equation}
Assume there exists a matrix $U = A^{-T}$ which diagonalizes both the covariance matrix of the between-class distribution $\Phi_b$ and the shared covariance matrix of the within-class distributions $\Phi_w$ of individual classes to $\Psi$ and $\mathcal{I}$, i.e. $U^T\Phi_bU=\Psi$ and $U^T\Phi_wU=\mathcal{I}$. Then, the simplified  model is written as:
\vspace{-1.5mm}
\begin{equation}
\begin{aligned}
x = m + Au\ \  where\ \  u\sim \mathcal{N}(v,\mathcal{I})\ \  and \ \  v \sim \mathcal{N}(0,\Psi)
\end{aligned}
\vspace{-1.5mm}
\label{e1}
\end{equation}
here, $x$ represents examples, i.e. speaker embedding; $v$ represents the class center; $u$ represents an example of that class in the projected space. Members of the same class share the class variable $v$, and the class-conditional distributions have a common covariance matrix $\mathcal{I}$. 

\subsubsection{ PLDA log-likelihood ratio}
Given two speaker embeddings, $x_1$ and $x_2$, PLDA provides the log-likelihood ratio between the same-speaker and different-speaker hypotheses, $\mathcal{H}_1$ and $\mathcal{H}_0$. The PLDA LLR is given by:
\begin{equation}
\begin{aligned}
LLR(x_1,x_2) = log\frac{p(x_1,x_2|\mathcal{H}_1)}{p(x_1,x_2|\mathcal{H}_0)}
\end{aligned}
\label{e2}
\end{equation}
According to Eq.\ref{e1}, embedding, i.e., $x$, is preprocessed by:
\begin{equation}
\begin{aligned}
u &= A^{-1}(x-m)\\
\end{aligned}
\end{equation}
Then, given the $\Psi$ and $\mathcal{I}$ are diagonal matrix,
the PLDA LLR can be expressed entirely in terms of scalar operations:

\begin{align}
&LLR(x_1,x_2)=log\dfrac{\int\mathcal{N}(u_1;v,I)\mathcal{N}(u_2;v,I)\mathcal{N}(v;0,\Psi)dv}{\mathcal{N}(u_1;0,\Psi+I)\mathcal{N}(u_2;0,\Psi+I)}\nonumber\\
&\quad\quad = \frac{1}{2}\sum_{i=1}^D\{c_i + m_i(2u_{1,i}u_{2,i} - \psi_i(u_{1,i}-u_{2,i})^2)\} \label{eq5}\\
&where:c_i = -log\frac{2\psi_i+1}{(\psi_i+1)^2} ,  m_i = \frac{\psi_i}{(2\psi_i+1)(\psi_i+1)}\nonumber\\
&Cos(x_1,x_2)=\sum_{i=1}^D{x_{1,i}x_{2,i}} \label{eq6}
\end{align}

By referring to the derivations, it is easy to find that PLDA LLR can be expressed as a weighted sum of weighted Cos similarity and Euclidean distance, which considers test data similarity and the distribution of the training data.

In same-domain situations, it is clear that larger $\psi_i$ is better because larger $\psi_i$ means a lower score when speakers are different by deriving expectations for each dimension in Eq.\ref{eqae}.
Furthermore, due to $u\sim \mathcal{N}(v,\mathcal{I})$ and $v \sim \mathcal{N}(0,\Psi)$, it is obvious that the larger the value of $\Psi$ , the better the PLDA discriminative capacity since within-class distribution is the identity matrix.
\begin{equation}
\begin{aligned}
E\{\frac{\psi_i}{(2\psi_i+1)(\psi_i+1)}(2u_{1,i}u_{2,i} - \psi_i(u_{1,i}-u_{2,i})^2\} \\=
\begin{cases}
\dfrac{\psi_i}{(2\psi_i+1)(\psi_i+1)}(2\psi_i-2\psi_i)=0 \ & \text{when} \quad \mathcal{H}_1\\
-\dfrac{2*\psi_i^2*(\psi_i + 1)}{(2\psi_i+1)(\psi_i+1)}\ & \text{when} \quad \mathcal{H}_0
\end{cases}
\end{aligned}
\label{eqae}
\end{equation}

In cross-domain situations, according to Eq.\ref{eq5}, i)the larger the testing data ($u_i$) variance, the better the discrimination, ii)the greater $\psi_i$, the heavier weight of the dimension in the score, iii)due to test data variance and PLDA variance are usually close, larger PLDA variance means more minor deviation, that is, more robust, because $m_i$ in Eq.\ref{eq5} is positively correlated with $1/\psi_i$ if $\psi_i>1$. Thus, PLDA may be more robust than Cos in cross-domain situations. The PLDA parameters $\{\mu,A,\Phi\}$ in the PLDA formula are optimal solutions in the training data space, thus, performance degradation is inevitable when encountering cross-domain problems unless the bias is the same in each dimension.

Here, one thing deserves to be stated. Compared to Cos, two major differences of PLDA are the underlying Gaussian assumption and usage of the training data distribution. Simple Gaussian assumptions of PLDA lead to desirable generalization but also limit the performance of DSEs. Utilizing the training data distribution improves the robustness but inevitably leads to performance degradation when encountering domain mismatch.

\subsection{Deep speaker embeddings}
As analyzed in the prior subsection, the distribution of embeddings has an influence on PLDA. 
Thus, we analyze the distribution of DSEs extracted from models with different (one or two) numbers of fully-connected layers in the segment-level part.
For convenience, embeddings extracted from the model with one \textit{fc} layer are denoted as \textit{1fc} embeddings, embeddings extracted from the first \textit{fc} layer, away from the classification layer, are denoted as \textit{2fc-1} embeddings, the second layer is denoted as \textit{2fc-2} embeddings.
\begin{figure}[!htbp]
\centering
\setlength{\belowcaptionskip}{0pt}
\vspace{-3mm}
\includegraphics[width=1.0\linewidth]{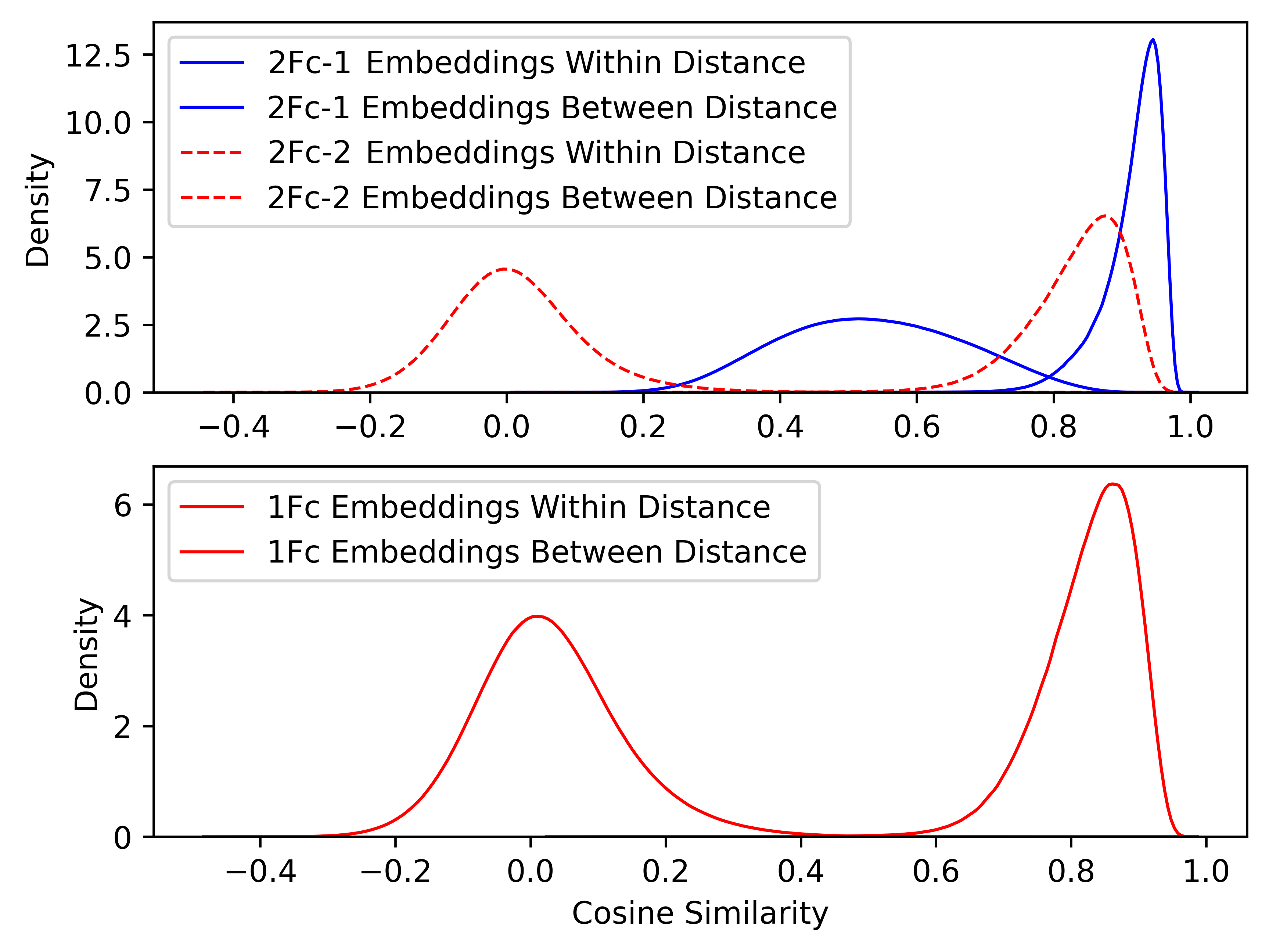}
\vspace{-7mm}
\caption{Within and between distances for embeddings of CTS superset extracted from ResNet34 with different number of segment-level layers}
\vspace{-4.5mm}
\label{f1}
\end{figure} 
As shown in Fig.\ref{f1}, the distribution of \textit{2fc-2} embeddings and \textit{1fc} embeddings are similar and are quite different from \textit{2fc-1} embeddings, but they all have strong discrimination.
Due to the compact distribution of \textit{2fc-1} embeddings, there is a significant overlap between intra- and inter-class distances. While no overlap exists for \textit{1fc}/\textit{2fc-2} embeddings due to the strong constraints of loss function, indicating the latter is better for Cos.  
Since there is just a simple non-linear relationship between \textit{2fc-1} embeddings and \textit{2fc-2} embeddings, \textit{2fc-1} embeddings have a stronger speaker discrimination potential. 
\begin{figure}[!htbp]
\centering
\setlength{\belowcaptionskip}{3pt}
\vspace{-2.5mm}
\includegraphics[width=1.0\linewidth]{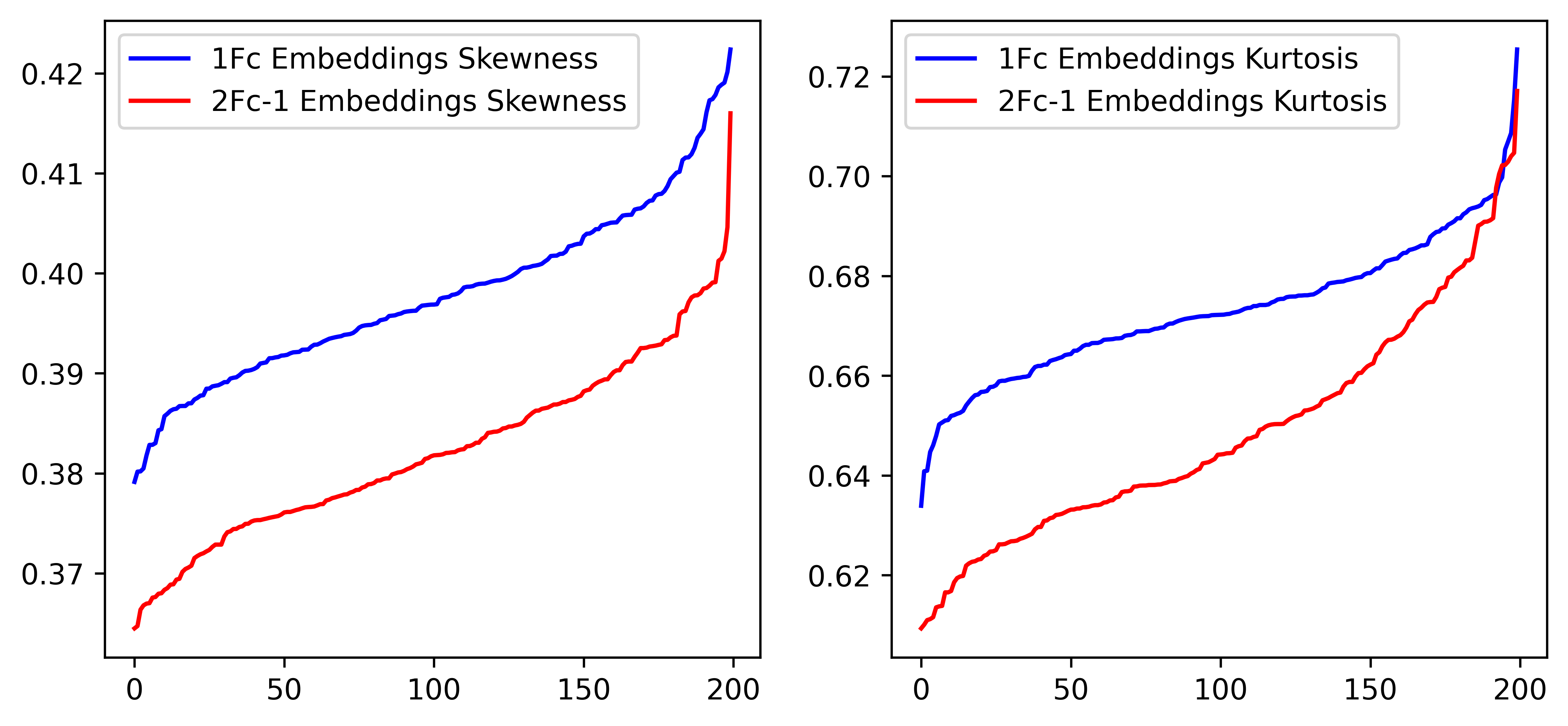}
\vspace{-6mm}
\caption{Skewness and kurtosis of the intraclass distribution of CTS set embeddings extracted from ResNet34 after LDA}
\vspace{-4mm}
\label{skew}
\end{figure} 

As depicted in Fig.\ref{skew}, we compute the skewness and kurtosis of the within-class distribution of embeddings after LDA for each dimension. Closer to zero are the two values, more Gaussian is a distribution. It is clear that the distributions of DSEs are non-gaussian, and the distribution of \textit{2fc-1} embeddings is closer to Gaussian than \textit{1fc}(or \textit{2fc-2}), showing that \textit{2fc-1} embeddings are better for PLDA.

\subsection{Back-ends selection for different situations }
\label{sec3.2}
Based on the analysis of the previous two subsections, we analyze the back-ends selection in different situations.

For the same-domain situations, since distributions are similar, the strong discrimination on the training set works perfectly on the test set, no matter what distribution they are. While the non-Gaussian distribution of DSEs causes great difficulty for PLDA and leads to a performance decrease. Study\cite{9023015} shows that the contribution of PLDA for deep speaker embeddings is regularization rather than discrimination, that is, PLDA tends to discover some underlying speech codes that are intrinsically Gaussian and comparable across speakers. Although these codes may be more generalized, they are more likely to result in performance degradation on same-domain test.
Thus, one conjecture can be made that \textit{1fc}/\textit{2fc-2}+Cos gets better performance in same-domain situations.

For the cross-domain situations, 
the strong discrimination of DSEs on the training set is not robust or even harmful, so the performance of Cos is usually poor. The more diverse the training set, the better the performance of Cos.
Compared to Cos, there are two advantages of PLDA. Firstly, the PLDA projected space constructed based on Gaussian assumptions is more generalizable. Secondly, more discriminative dimensions in the space are more robust to cross-domain.
An important measure of cross-domain is whether the test data variance deviates from the training data in the PLDA projected space. Usually, the trend in variance with dimensions is similar for most domain data. Sec4.4 will present it in detail.
Then, dimensions with larger PLDA variance, i.e. more discriminative, are more insensitive to variance variability between PLDA and test data to be more robust to cross-domain data. 
Therefore, the other conjecture can be made that \textit{2fc-1}+PLDA has a better performance in cross-domain situations.

\section{Experiments Setting}
We conduct experiments on four conditions, single/multi-domain training and same/cross-domain test.
Model training is conducted on three training sets, \textbf{{(a)}} the VoxCeleb2\cite{vox1,vox2} dev part, which contains speech from 5994 speakers. The experiments settings are same as \cite{xiao21b_interspeech}. No speech augmentation and voice activity detection (VAD) are used in the series of experiments. \textbf{{(b)}} NIST SRE CTS superset (CTS)\cite{cts}, contains 6867 speakers. 81-dimensions fbanks spanning the frequency range 40-3800Hz are used and 3-dimensions pitch features are concatenated. All data are augmented by convolving with far-field Room Impulse Responses (RIRs) and adding noise from the MUSAN corpus. An energy-based and harmonics-based VAD is used to drop the non-speech frames.
\textbf{{(c)}} CTS and Vox2Cat, which contain a total of 12861 speakers. Due to the short duration of speech from VoxCeleb2, we concatenate the subsegments belonging to the same original video into a unique segment, named as Vox2Cat. All audios are converted to 8kHz-16bit-PCM in WAV format files. Speech augmentation and VAD are the same as \textbf{{(b)}}.

We evaluate our models on two series of sets, (i) Vox1-O/H, (ii)SRE21-dev\&eval. The former test sets are well known, and we briefly introduce the latter, SRE21-dev\&eval. 
Compared to the prior SREs, the key challenges of SRE21 are multi-channel and multi-language speaker recognition based on audio-from video and telephone speech segments, which cause extremely serious enrollment-test and training-test cross-domain problems. The SRE21 sets we used in all experiments are preprocessed by codec and denoise.

We conduct the experiments on two models, ResNet34 and ETDNN. ASP layer and circle loss\cite{xiao21b_interspeech} are used in all models. All models are trained with stochastic gradient descent and random chunk size. \textbf{{(a)}}-models interval is set to [200,400],[300-500] and [400,600], \textbf{{(b)}}- and \textbf{{(c)}}-models interval is set to  [400,800], [600-1000], [800-1000/1200] in three training stages.
The performance on the Vox series test set is gauged in terms of the EER, minDCF with $p_{target}=0.01$, and the NIST SRE21 sets are gauged with EER and minimum $C_{primary}$, which are calculated with the default scripts\cite{sadjadi2021nist}.

\section{Results and Analysis}
\vspace{-0.5mm}
In this section, we explore how to select back-ends in various situations by experiments, noting that Cos of \textit{2fc} uses \textit{2fc-2} embeddings, PLDA uses \textit{2fc-1} embeddings in all Tables\footnote{The performance of Cos using \textit{2fc-1} embeddings is unacceptable due to its compact distribution, and we do not present it.
Also, since the distribution and PLDA scoring of \textit{2fc-1} embeddings is similar to \textit{1fc} embeddings, we also overlook it.}. 
\vspace{-1.0mm}
\subsection{single-domain training, same-domain test}
\vspace{-0.5mm}
\label{sec4.1}
We conduct the same-domain experiments on the VoxCeleb sets with ResNet34 and ETDNN with 512 channels.

\begin{table}[htbp]
\vspace{-2.5mm}
  \centering
  \caption{Results of Vox test sets when training data is \textbf{Vox2}}
\vspace{-3mm}
\setlength{\tabcolsep}{1.4mm}{
    \begin{tabular}{ccc|cccc}
    \toprule
    \multicolumn{3}{c|}{\multirow{2}[2]{*}{}} & \multicolumn{2}{c}{Vox1-O} & \multicolumn{2}{c}{Vox1-H} \\
    \multicolumn{3}{c|}{} & EER   & minDCF & EER   & minDCF \\
\cmidrule{1-7}    \multicolumn{1}{c}{\multirow{5}[6]{*}{ResNet34}} & \multicolumn{2}{c|}{\cite{ecapa}} & 1.19  & 0.159 & 2.46  & 0.229 \\
\cmidrule{2-7}          & \multirow{2}[2]{*}{\textit{1fc}} & Cos   & \textbf{1.39} & \textbf{0.115} & \textbf{2.56} & \textbf{0.237} \\
          &       & PLDA  & 2.24  & 0.196 & 3.82  & 0.367 \\
\cmidrule{2-7}          & \multirow{2}[2]{*}{\textit{2fc}} & Cos   & 1.92  & 0.191 & 3.28  & 0.287 \\
          &       & PLDA  & 1.45  & 0.182 & 2.67  & 0.260 \\
    \midrule
    \multicolumn{1}{c}{\multirow{5}[6]{*}{ETDNN}} & \multicolumn{2}{c|}{\cite{ecapa}} & 1.49  & 0.161 & 2.69  & 0.242 \\
\cmidrule{2-7}          & \multirow{2}[2]{*}{\textit{1fc}} & Cos   & \textbf{1.62} & \textbf{0.179} & \textbf{2.88} & \textbf{0.265} \\
          &       & PLDA  & 2.21  & 0.263 & 3.46  & 0.328 \\
\cmidrule{2-7}          & \multirow{2}[2]{*}{\textit{2fc}} & Cos   & 2.1   & 0.191 & 3.23  & 0.292 \\
          &       & PLDA  & 1.99  & 0.213 & 3.46  & 0.332 \\
    \bottomrule
	\end{tabular}
  \label{t1}}%
\vspace{-3.5mm}
\end{table}%

As shown in Table~\ref{t1}, three things can be observed: i) severe performance degradation is caused by adopting PLDA scoring on \textit{1fc} embeddings, compared to Cos, ii) PLDA scoring of \textit{2fc-1} embeddings is sightly better than Cos of \textit{2fc-2} embeddings, iii)the performance of Cos using \textit{1fc} embeddings is better than the performance of PLDA scoring using \textit{2fc-1} embeddings.
\setcounter{table}{2}
\begin{table*}[htbp]
\vspace{-4mm}
  \centering
  \caption{Results of SRE21\&Vox test sets when training data is \textbf{CTS and Vox2Cat}}
	\vspace{-2.5mm}
    \begin{tabular}{ccc|cccc|c|cccc}
    \toprule
    \multicolumn{3}{c|}{\multirow{2}[2]{*}{}} & \multicolumn{2}{c}{sre21-dev} & \multicolumn{2}{c|}{sre21-eval} & \multirow{2}[2]{*}{} & \multicolumn{2}{c}{vox1-O} & \multicolumn{2}{c}{Vox1-H} \\
    \multicolumn{3}{c|}{} & EER   & minCp & EER   & minCp &       & EER   & mindcf & EER   & mindcf \\
    \midrule
    \multirow{4}[4]{*}{ResNet34} & \multirow{2}[2]{*}{\textit{1fc}} & Cos   & 8.39  & 0.589 & 8.47  & 0.506 & Cos   & \textbf{2.15} & \textbf{0.247} & \textbf{3.63} & \textbf{0.312} \\
          &       & CTS-PLDA & 7.40   & 0.634 & 8.29  & 0.632 & Vox2-PLDA & 1.94  & 0.294 & 3.86  & 0.386 \\
\cmidrule{2-12}          & \multirow{2}[2]{*}{\textit{2fc}} & Cos   & 9.76  & 0.585 & 8.88  & 0.497 & Cos   & 2.04  & 0.284 & 3.69  & 0.323 \\
          &       & CTS-PLDA & \textbf{6.15} & \textbf{0.395} & \textbf{6.57} & \textbf{0.382} & Vox2-PLDA & 2.14  & 0.31  & 4.24  & 0.421 \\
    \midrule
    \multirow{4}[4]{*}{ETDNN} & \multirow{2}[2]{*}{\textit{1fc}} & Cos   & 7.98  & 0.488 & 7.48  & 0.481 & Cos   & \textbf{2.08} & \textbf{0.258} & \textbf{3.57} & \textbf{0.305} \\
          &       & CTS-PLDA & 7.49  & 0.537 & 7.77  & 0.547 & Vox2-PLDA & 2.48  & 0.308 & 4.64  & 0.445 \\
\cmidrule{2-12}          & \multirow{2}[2]{*}{\textit{2fc}} & Cos   & 8.49  & 0.563 & 8.19  & 0.505 & Cos   & 2.47  & 0.295 & 4.07  & 0.345 \\
          &       & CTS-PLDA & \textbf{7.05} & \textbf{0.445} & \textbf{7.37} & \textbf{0.502} & Vox2-PLDA & 2.53  & 0.319 & 4.68  & 0.464 \\
    \bottomrule
    \end{tabular}%
  \label{t3}%
\vspace{-5mm}
\end{table*}%

Thus, one rough conclusion can be drawn, in \textit{single-domain training, same-domain test} conditions, the performance of Cos is better than PLDA. Considering simplicity and performance, Cos is a better choice. These things prove our analysis and conjectures.
The reason why the performance of Cos of \textit{2fc-2} embeddings is worse than \textit{1fc} is information loss caused by ReLU activation functions. Using other activation functions partly alleviates degradation and gets better Cos performance but slightly damages PLDA scoring of \textit{2fc-1} embeddings.

\vspace{-1mm}
\subsection{Single-domain training, cross-domain test}
\label{sc}
we conduct experiments on the CTS with ResNet34 and ETDNN with 1024 channels. Models are evaluated on SRE21 dev\&eval. Results are shown in Table~\ref{t2} and no adaption is applied. Some different things happen compared to Sec\ref{sec4.1}.  
\setcounter{table}{1}
\begin{table}[!th]
  \centering
  \caption{Results of SRE21 sets when training data is \textbf{CTS}}
	\vspace{-2.5mm}
\setlength{\tabcolsep}{1.4mm}{
    \begin{tabular}{ccc|cccc}
    \toprule
    \multicolumn{2}{c}{\multirow{2}[2]{*}{}} & \multirow{2}[2]{*}{} & \multicolumn{2}{c}{SRE21-dev} & \multicolumn{2}{c}{SRE21-eval} \\
    \multicolumn{2}{c}{} &       & EER   & minCp & EER   & minCp \\
    \midrule
    \multirow{4}[4]{*}{ResNet34} & \multirow{2}[2]{*}{\textit{1fc}} & Cos   & 9.94  & 0.644 & 9.74  & 0.548 \\
          &       & PLDA  & 6.87  & 0.446 & 6.76  & 0.471 \\
\cmidrule{2-7}          & \multirow{2}[2]{*}{\textit{2fc}} & Cos   & 13.27 & 0.719 & 12.27 & 0.621 \\
          &       & PLDA  & \textbf{6.58} & \textbf{0.423} & \textbf{6.97} & \textbf{0.434} \\
    \midrule
    \multirow{4}[4]{*}{ETDNN} & \multirow{2}[2]{*}{\textit{1fc}} & Cos   & 10.00    & 0.586 & 9.65  & 0.553 \\
          &       & PLDA  & 9.26  & 0.576 & 8.59  & 0.532 \\
\cmidrule{2-7}          & \multirow{2}[2]{*}{\textit{2fc}} & Cos   & 9.90   & 0.586 & 9.60   & 0.548 \\
          &       & PLDA  & \textbf{10.09} & \textbf{0.468} & \textbf{8.86} & \textbf{0.474} \\
    \bottomrule
    \end{tabular}}%
  \label{t2}%
\vspace{-6mm}
\end{table}%

Firstly, PLDA scoring is significantly better than Cos in all conditions, about 4\%-15\% in \textit{1fc} and about 15\%-30\% in \textit{2fc-1} in term of min$C_p$. A larger improvement of PLDA in \textit{2fc-1} further proves our conjecture in Sec\ref{sec3.2}.
Secondly, although Cos of \textit{1fc} is better than \textit{2fc-1}, the performance of PLDA scoring using \textit{2fc-1} embeddings is sightly better than the performance of PLDA scoring using \textit{1fc} embeddings, and is the best compared to others in SRE21, about 8\%-10\% on min$C_p$.
Thus, a new conclusion could be drawn, in \textit{single-domain training, cross-domain test}, PLDA is a better selection.

\subsection{Multi-domain training}

Results are presented in Table~\ref{t3}, PLDA for Vox sets is trained by Vox2 while that for SRE21 is trained by CTS. Three things worth noting: i) Although \textit{2fc-1}+PLDA is still the best in cross-domain situations, the improvements by PLDA is fewer than Sec\ref{sc}, and the performance of Cos using \textit{1fc} and \textit{2fc-2} is better than Sec\ref{sc}. 
ii) Different from sec~\ref{sc}, sightly performance degradation is caused by PLDA for \textit{1fc}, and even get worse results than sec~\ref{sc}. This is puzzling and we will analyze it in later work.
iii) PLDA causes universal performance degradation in the same-domain test, either in \textit{1fc} or \textit{2fc}. 
Thus far, we conclude that Cos with \textit{1fc} is better in same-domain situations and PLDA with \textit{2fc-1} is better in cross-domain situations.

\subsection{Domain adaption analysis}

As discussed in Sec\ref{sec211}, differences between PLDA variance and test data variance reflects the domain mismatch.
\setcounter{table}{3}
 \begin{table}[htbp]
  \centering
	\vspace{-2mm}
  \caption{Results of SRE21 after adaption when training data is \textbf{CTS and Vox2cat}}
	\vspace{-3mm}
    \begin{tabular}{cc|cccc}
    \toprule
    \multicolumn{2}{c|}{\multirow{2}[2]{*}{}} & \multicolumn{2}{c}{SRE21-dev} & \multicolumn{2}{c}{SRE21-eval} \\
    \multicolumn{2}{c|}{} & EER   & minCp & EER   & minCp \\
    \midrule
    \multirow{2}[1]{*}{ResNet34} & coral & 8.72  & 0.418 & 7.33  & 0.432 \\
          & coral+ & \textbf{5.88} & \textbf{0.321} & \textbf{4.99} & \textbf{0.330} \\
    \multirow{2}[1]{*}{ETDNN} & coral & 9.91  & 0.459 & 8.34  & 0.498 \\
          & coral+ & \textbf{6.15} & \textbf{0.364} & \textbf{5.82} & \textbf{0.394} \\
    \midrule
    \multicolumn{2}{c|}{ResNet34\cite{avdeeva2021stc}} & 6.10   & 0.377 & —     & — \\
    \multicolumn{2}{c|}{ECAPA\cite{avdeeva2021stc}} & 5.94  & 0.352 & —     & — \\
    \bottomrule
    \end{tabular}%
  \label{t5}%
\vspace{-3mm}
\end{table}%
\begin{figure}[!bh]
\vspace{-0mm}
\centering
\setlength{\belowcaptionskip}{3pt}
\includegraphics[width=0.86\linewidth]{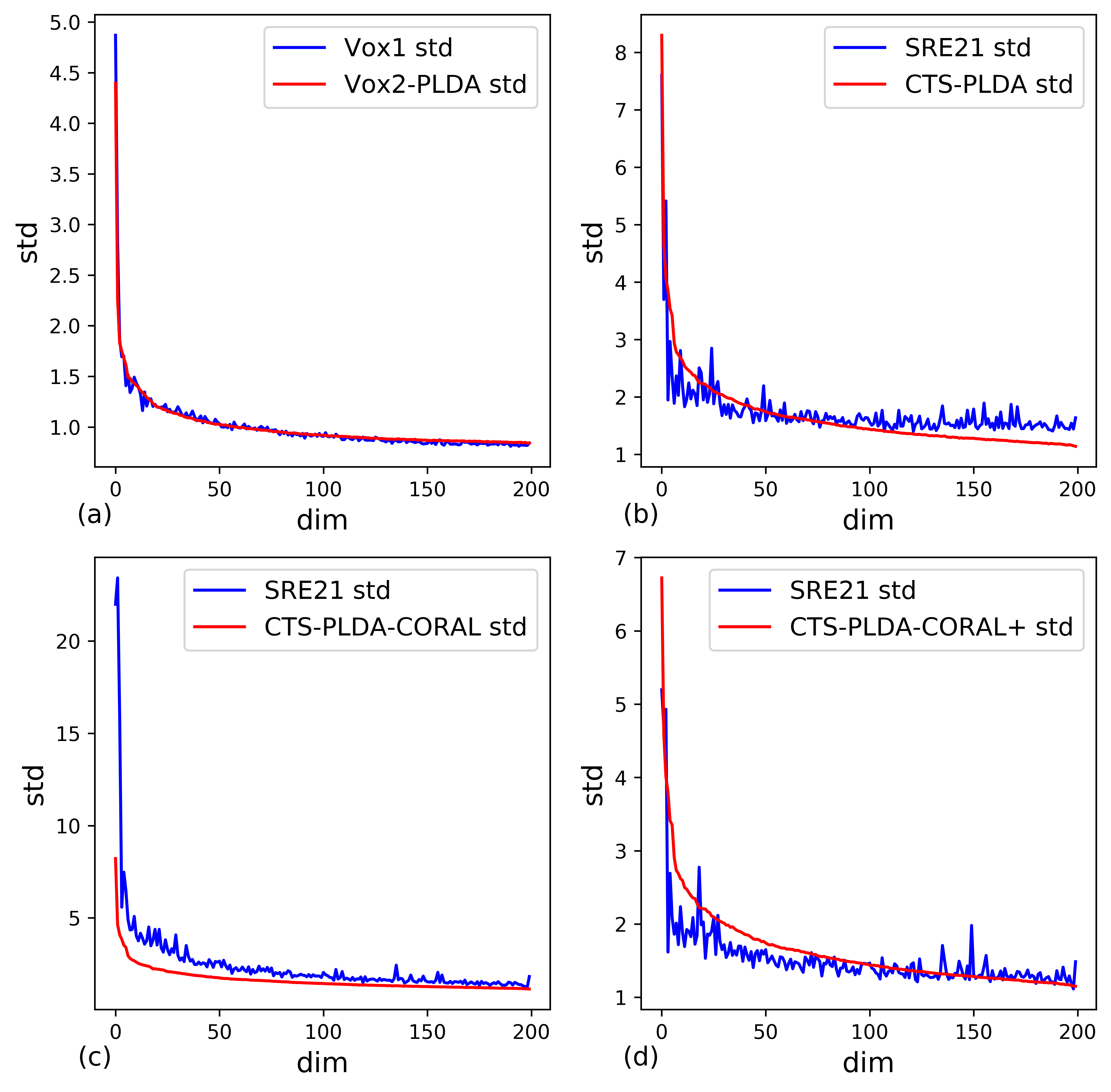}
\vspace{-3mm}
\caption{Std of test data and PLDA for ResNet34}
\label{embed}
\vspace{-5.0mm}
\end{figure} 
As seen in Fig.\ref{embed} (a)\&(b), the variance of Vox1 sets and Vox2 training data are approximately consistent, while SRE21 sets and CTS training sets are quite different, especially at lower variance. The standard deviation is used in Fig. \ref{embed} to enlarge differences.

Thus, the main role of back-ends adaption methods is to align training-test statistics, mean and variance.
The CORAL algorithm\cite{sun2016return}, which is simple, focus on aligning the covariance between different domain embeddings by whitening and re-colouring. They can be equivalently applied to back-ends since operations are linear.
A major issue of the CORAL algorithm is that it is overconfidence in in-domain data distribution, especially with limited in-domain data. Considering these, the CORAL+ algorithm introduces interpolation and regularization. Since larger PLDA variance means less deviation, as analyzed in Sec\ref{sec211}. Thus, one $max()$ operator is introduced in the CORAL+ algorithm\cite{lee2019coral+}.
Fig.\ref{embed} (c)\&(d) shows the variance after CORAL and CORAL+. It is found that the lower variances are aligned by CORAL+, while CORAL does not work. PLDA results after adaption are displayed in Table~\ref{t5}. The CORAL+ achieve about 15\%-20\% improvements on $minC_p$.

\section{Conclusions}

The paper systematically explores back-ends selection for DSEs in different situations by theory analysis and experiments validation. By analyzing PLDA and the properties of DSEs, we conjecture that Cos is better in same-domain situations and PLDA is better in cross-domain situations. Then, we validate our conjecture by conducting experiments on VoxCeleb and CTS. 
Differences between the variances of PLDA and test data cause performance degradation in cross-domain situations.

\bibliographystyle{IEEEtran}

\bibliography{mybib}

\begin{thebibliography}{10}
\providecommand{\url}[1]{#1}
\csname url@samestyle\endcsname
\providecommand{\newblock}{\relax}
\providecommand{\bibinfo}[2]{#2}
\providecommand{\BIBentrySTDinterwordspacing}{\spaceskip=0pt\relax}
\providecommand{\BIBentryALTinterwordstretchfactor}{4}
\providecommand{\BIBentryALTinterwordspacing}{\spaceskip=\fontdimen2\font plus
\BIBentryALTinterwordstretchfactor\fontdimen3\font minus
  \fontdimen4\font\relax}
\providecommand{\BIBforeignlanguage}[2]{{%
\expandafter\ifx\csname l@#1\endcsname\relax
\typeout{** WARNING: IEEEtran.bst: No hyphenation pattern has been}%
\typeout{** loaded for the language `#1'. Using the pattern for}%
\typeout{** the default language instead.}%
\else
\language=\csname l@#1\endcsname
\fi
#2}}
\providecommand{\BIBdecl}{\relax}
\BIBdecl

\bibitem{ivector}
N.~Dehak, P.~J. Kenny, R.~Dehak, P.~Dumouchel, and P.~Ouellet, ``Front-end
  factor analysis for speaker verification,'' \emph{IEEE Transactions on Audio,
  Speech, and Language Processing}, vol.~19, no.~4, pp. 788--798, 2010.

\bibitem{plda}
S.~Ioffe, ``Probabilistic linear discriminant analysis,'' in \emph{European
  Conference on Computer Vision}.\hskip 1em plus 0.5em minus 0.4em\relax
  Springer, 2006, pp. 531--542.

\bibitem{plda02}
P.~Kenny, ``Bayesian speaker verification with, heavy tailed priors,''
  \emph{Proc. Odyssey 2010}, 2010.

\bibitem{xvectorbefore}
D.~Snyder, D.~Garcia-Romero, D.~Povey, and S.~Khudanpur, ``Deep neural network
  embeddings for text-independent speaker verification,'' in \emph{Proc.
  Interspeech 2017}, 2017, pp. 999--1003.

\bibitem{snyder2019speaker}
D.~Snyder, D.~Garcia-Romero, G.~Sell, A.~McCree, D.~Povey, and S.~Khudanpur,
  ``Speaker recognition for multi-speaker conversations using x-vectors,'' in
  \emph{ICASSP 2019-2019 IEEE International conference on acoustics, speech and
  signal processing (ICASSP)}.\hskip 1em plus 0.5em minus 0.4em\relax IEEE,
  2019, pp. 5796--5800.

\bibitem{liming}
W.~Cai, J.~Chen, and M.~Li, ``Exploring the encoding layer and loss function in
  end-to-end speaker and language recognition system,'' in \emph{Proc. Odyssey
  2018 The Speaker and Language Recognition Workshop}, 2018, pp. 74--81.

\bibitem{resnet02}
N.~Li, D.~Tuo, D.~Su, Z.~Li, D.~Yu, and A.~Tencent, ``Deep discriminative
  embeddings for duration robust speaker verification.'' in \emph{Interspeech},
  2018, pp. 2262--2266.

\bibitem{res2net}
S.~{Gao}, M.~M. {Cheng}, K.~{Zhao}, X.~Y. {Zhang}, M.~H. {Yang}, and P.~H.~S.
  {Torr}, ``Res2net: A new multi-scale backbone architecture,'' \emph{IEEE
  Transactions on Pattern Analysis and Machine Intelligence}, pp. 1--1, 2019.

\bibitem{ecapa}
B.~Desplanques, J.~Thienpondt, and K.~Demuynck, ``Ecapa-tdnn: Emphasized
  channel attention, propagation and aggregation in tdnn based speaker
  verification,'' \emph{Proc. Interspeech 2020}, pp. 3830--3834, 2020.

\bibitem{okabe2018attentive}
K.~Okabe, T.~Koshinaka, and K.~Shinoda, ``Attentive statistics pooling for deep
  speaker embedding,'' \emph{Proc. Interspeech 2018}, pp. 2252--2256, 2018.

\bibitem{tang2019deep}
Y.~Tang, G.~Ding, J.~Huang, X.~He, and B.~Zhou, ``Deep speaker embedding
  learning with multi-level pooling for text-independent speaker
  verification,'' in \emph{ICASSP 2019-2019 IEEE International Conference on
  Acoustics, Speech and Signal Processing (ICASSP)}.\hskip 1em plus 0.5em minus
  0.4em\relax IEEE, 2019, pp. 6116--6120.

\bibitem{wang2020multi}
Z.~Wang, K.~Yao, X.~Li, and S.~Fang, ``Multi-resolution multi-head attention in
  deep speaker embedding,'' in \emph{ICASSP 2020-2020 IEEE International
  Conference on Acoustics, Speech and Signal Processing (ICASSP)}.\hskip 1em
  plus 0.5em minus 0.4em\relax IEEE, 2020, pp. 6464--6468.

\bibitem{liu2019large}
Y.~Liu, L.~He, and J.~Liu, ``Large margin softmax loss for speaker
  verification,'' \emph{arXiv preprint arXiv:1904.03479}, 2019.

\bibitem{zhou2020dynamic}
D.~Zhou, L.~Wang, K.~A. Lee, Y.~Wu, M.~Liu, J.~Dang, and J.~Wei, ``Dynamic
  margin softmax loss for speaker verification,'' \emph{Proc. Interspeech
  2020}, pp. 3800--3804, 2020.

\bibitem{8759958}
S.~Wang, Z.~Huang, Y.~Qian, and K.~Yu, ``Discriminative neural embedding
  learning for short-duration text-independent speaker verification,''
  \emph{IEEE/ACM Transactions on Audio, Speech, and Language Processing},
  vol.~27, no.~11, pp. 1686--1696, 2019.

\bibitem{xiao21b_interspeech}
R.~Xiao, X.~Miao, W.~Wang, P.~Zhang, B.~Cai, and L.~Luo, ``{Adaptive Margin
  Circle Loss for Speaker Verification},'' in \emph{Proc. Interspeech 2021},
  2021, pp. 4618--4622.

\bibitem{1fcetdnn}
D.~Garcia-Romero, A.~McCree, D.~Snyder, and G.~Sell, ``Jhu-hltcoe system for
  the voxsrc speaker recognition challenge,'' in \emph{ICASSP 2020 - 2020 IEEE
  International Conference on Acoustics, Speech and Signal Processing}, 2020,
  pp. 7559--7563.

\bibitem{thienpondt2020idlab}
J.~Thienpondt, B.~Desplanques, and K.~Demuynck, ``The idlab voxceleb speaker
  recognition challenge 2020 system description,'' \emph{arXiv preprint
  arXiv:2010.12468}, 2020.

\bibitem{wang2020dku}
W.~Wang, D.~Cai, X.~Qin, and M.~Li, ``The dku-dukeece systems for voxceleb
  speaker recognition challenge 2020,'' \emph{arXiv preprint arXiv:2010.12731},
  2020.

\bibitem{zhang2021beijing}
L.~Zhang, H.~Zhao, Q.~Meng, Y.~Chen, M.~Liu, and L.~Xie, ``Beijing zkj-npu
  speaker verification system for voxceleb speaker recognition challenge
  2021,'' \emph{arXiv preprint arXiv:2109.03568}, 2021.

\bibitem{brummerbut}
N.~Brummer, L.~Burget, O.~Glembek, P.~Matejka, L.~Mo{\v{s}}ner, O.~Novotn{\`y},
  O.~Plchot, J.~Rohdin, A.~Silnova, T.~Stafylakis \emph{et~al.}, ``But+ omilia
  system description voxceleb speaker recognition challenge 2020.''

\bibitem{wang2021xmuspeech}
J.~Wang, F.~Tong, Z.~Chen, L.~Li, Q.~Hong, and H.~Zhou, ``Xmuspeech system for
  voxceleb speaker recognition challenge 2021,'' \emph{arXiv preprint
  arXiv:2109.02549}, 2021.

\bibitem{lee2019i4u}
K.~A. Lee, V.~Hautamaki, T.~Kinnunen, H.~Yamamoto, K.~Okabe, V.~Vestman,
  J.~Huang, G.~Ding, H.~Sun, A.~Larcher \emph{et~al.}, ``I4u submission to nist
  sre 2018: Leveraging from a decade of shared experiences,'' \emph{arXiv
  preprint arXiv:1904.07386}, 2019.

\bibitem{lee2020nec}
K.~A. Lee, K.~Okabe, H.~Yamamoto, Q.~Wang, L.~Guo, T.~Koshinaka, J.~Zhang,
  K.~Ishikawa, and K.~Shinoda, ``Nec-tt speaker verification system for sre'19
  cts challenge.'' in \emph{INTERSPEECH}, 2020, pp. 2227--2231.

\bibitem{villalba2019state}
J.~Villalba, N.~Chen, D.~Snyder, D.~Garcia-Romero, A.~McCree, G.~Sell,
  J.~Borgstrom, F.~Richardson, S.~Shon, F.~Grondin \emph{et~al.},
  ``State-of-the-art speaker recognition for telephone and video speech: the
  jhu-mit submission for nist sre18,'' 2019.

\bibitem{li2020thuee}
R.~Li, T.~Liang, D.~Song, Y.~Liu, Y.~Wu, C.~Xu, P.~Ouyang, X.~Zhang, X.~Chen,
  W.~Zhang \emph{et~al.}, ``Thuee system for nist sre19 cts challenge.'' in
  \emph{INTERSPEECH}, 2020, pp. 2232--2236.

\bibitem{li2020voiceai}
R.~Li, D.~Chen, and W.~Zhang, ``Voiceai systems to nist sre19 evaluation:
  Robust speaker recognition on conversational telephone speech,'' in
  \emph{ICASSP 2020-2020 IEEE International Conference on Acoustics, Speech and
  Signal Processing (ICASSP)}.\hskip 1em plus 0.5em minus 0.4em\relax IEEE,
  2020, pp. 6459--6463.

\bibitem{alam2019abc}
J.~Alam, G.~Boulianne, O.~Glembek, A.~D. Lozano, P.~Matejka, P.~Mizera,
  J.~Monteiro, L.~Mo{\v{s}}ner, O.~Novotn{\`y}, O.~Plchot \emph{et~al.}, ``Abc
  nist sre 2019 cts system description,'' \emph{Proceedings of NIST}, pp. 1--6,
  2019.

\bibitem{ramoji2020leap}
S.~Ramoji, P.~Krishnan, B.~Mysore, P.~Singh, and S.~Ganapathy, ``Leap system
  for sre 2019 cts challenge-improvements and error analysis,'' in \emph{Proc.
  Odyssey 2020 The Speaker and Language Recognition Workshop}, 2020, pp.
  281--288.

\bibitem{sadjadi2021nist}
O.~Sadjadi, C.~Greenberg, E.~Singer, L.~Mason, and D.~Reynolds, ``Nist 2021
  speaker recognition evaluation plan,'' 2021.

\bibitem{avdeeva2021stc}
A.~Avdeeva, A.~Gusev, I.~Korsunov, A.~Kozlov, G.~Lavrentyeva, S.~Novoselov,
  T.~Pekhovsky, A.~Shulipa, A.~Vinogradova, V.~Volokhov \emph{et~al.}, ``Stc
  speaker recognition systems for the nist sre 2021,'' \emph{arXiv preprint
  arXiv:2111.02298}, 2021.

\bibitem{9023015}
X.~Wang, L.~Li, and D.~Wang, ``Vae-based domain adaptation for speaker
  verification,'' in \emph{2019 Asia-Pacific Signal and Information Processing
  Association Annual Summit and Conference (APSIPA ASC)}, 2019, pp. 535--539.

\bibitem{9296778}
Y.~Cai, L.~Li, A.~Abel, X.~Zhu, and D.~Wang, ``Deep normalization for speaker
  vectors,'' \emph{IEEE/ACM Transactions on Audio, Speech, and Language
  Processing}, vol.~29, pp. 733--744, 2021.

\bibitem{vox1}
A.~Nagrani, J.~S. Chung, and A.~Zisserman, ``Voxceleb: a large-scale speaker
  identification dataset,'' \emph{arXiv preprint arXiv:1706.08612}, 2017.

\bibitem{vox2}
J.~S. Chung, A.~Nagrani, and A.~Zisserman, ``{VoxCeleb2: Deep Speaker
  Recognition},'' in \emph{Proc. Interspeech 2018}, 2018, pp. 1086--1090.

\bibitem{cts}
S.~O. Sadjadi, ``Nist sre cts superset: A large-scale dataset for telephony
  speaker recognition,'' \emph{arXiv preprint arXiv:2108.07118}, 2021.

\bibitem{sun2016return}
B.~Sun, J.~Feng, and K.~Saenko, ``Return of frustratingly easy domain
  adaptation,'' in \emph{Proceedings of the AAAI Conference on Artificial
  Intelligence}, vol.~30, no.~1, 2016.

\bibitem{lee2019coral+}
K.~A. Lee, Q.~Wang, and T.~Koshinaka, ``The coral+ algorithm for unsupervised
  domain adaptation of plda,'' in \emph{ICASSP 2019-2019 IEEE International
  Conference on Acoustics, Speech and Signal Processing (ICASSP)}.\hskip 1em
  plus 0.5em minus 0.4em\relax IEEE, 2019, pp. 5821--5825.

\end{thebibliography}

\end{document}